\documentclass[12pt]{iopart}
\usepackage[english]{babel}
\usepackage{indentfirst}
\usepackage{graphicx}
\usepackage{subfigure}
\usepackage{dcolumn}
\usepackage{bm}
\usepackage{color}
\begin{document}

\title{One-dimensional multicomponent Fermi gas in a trap: quantum Monte Carlo study}
\author{N. Matveeva}
\address{Universite Grenoble Alpes, CNRS, LPMMC, UMR 5493, F-38042 Grenoble, France}
\author{G. E. Astrakharchik}
\address{Departament de F\'{\i}sica, Universitat Polit\`{e}cnica de Catalunya, 08034 Barcelona, Spain}
\date{\date}

\begin{abstract}
One-dimensional world is very unusual as there is an interplay between quantum statistics and geometry, and a strong short-range repulsion between atoms mimics Fermi exclusion principle, fermionizing the system.
Instead, a system with a large number of components with a single atom in each, on the opposite acquires many bosonic properties.
We study the ground-state properties a multi-component Fermi gas trapped in a harmonic trap by fixed-node diffusion Monte Carlo method.
We investigate how the energetic properties (energy, contact) and correlation functions (density profile and momentum distribution) evolve as the number of components is changed.
It is shown that the system fermionizes in the limit of strong interactions.
Analytical expression are derived in the limit of weak interactions within the local density approximation for arbitrary number of components and for one plus one particle using an exact solution.
\end{abstract}

\maketitle

\section{Introduction}

Quantum one-dimensional systems can be realized in ultracold gases confined in cigar-shaped traps\cite{Greiner01,Moritz03,Kinoshita04,Haller09,Fang14,Guan13}.
At ultracold temperatures atoms manifest different behavior, depending if they obey Fermi-Dirac or Bose-Einstein statistics.
In terms of the wave function, a different symmetry is realized with the respect to an exchange of two particles, bosons having the symmetric wave function and fermions the antisymmetric one.
A peculiarity of one dimension is that reduced geometry imposes certain limitations on probing the symmetry due to an exchange.
The only way of exchanging two particles on a line is to move one particle though the other.
This leads to important consequences when particles interact via an infinite repulsion.
In this case, known as Tonks-Girardeau limit, systems might acquire some fermionic properties.
The energy of a homogeneous Bose gas with contact interaction (Lieb-Liniger model) can be exactly found\cite{Lieb63} using Bethe {\it ansatz} method and follows a crossover from mean-field Gross-Pitaevskii gas to Tonks-Girardeau gas, in which the energetic properties are the same as for ideal Fermions\cite{Kinoshita04,Kinoshita05,Paredes04}.
The equation of state can be probed\cite{Moritz03,Haller09,Fang14} by exciting the breathing mode in a trapped gas, as the frequency of collective oscillations depends on the compressibility.
This allowed to observe a smooth crossover from the mean-field Gross-Pitaevskii to the Tonks-Girardeau/ideal Fermi gas value\cite{Haller09}.
For a large number of particles, the breathing mode frequency in the crossover is well described\cite{Menotti02} within the local density approximation (LDA) which relies on the knowledge of the homogeneous equation of state\cite{Lieb63}.
Instead for a small number of particles, the LDA approach misses the reentrant behavior in the Gross-Pitaevskii -- Gaussian crossover, which first was observed experimentally\cite{Haller09} and later quantitatively explained\cite{Gudyma15} using quantum Monte Carlo method.

The interplay between repulsive interactions and statistics was clearly demonstrated in a few-atom experiments in Selim Jochim's group\cite{Zurn12} where it was shown that two component fermions with a single atom in each component {\em fermionize} when the interaction between them becomes infinite.
The question becomes more elaborate when the number of components becomes large.
A system of $N_c$ component fermions with a single atom in each component should have a similar energetic properties as a single component Bose gas consisting of $N_c$ atoms\cite{Yang11}.
Recently a six-component mixture of one-dimensional fermions was realized in LENS group\cite{Pagano14}.
The measured frequency of the breathing mode approaches to that of a Bose system as $N_c$ is increased from 1 to 6.
It was observed that the momentum distribution increases its width as the number of components is increased, keeping the number of atoms in each spin component fixed.

The Bethe {\it ansatz} theory is well suited for finding the energetic properties in a homogeneous geometry, predicting the equation of state for the case of $N_c=2$ components \cite{Yang67,Krivnov75} and an arbitrary $N_c$\cite{LiuDrummond08,Guan12}.
Unfortunately Bethe {\it ansatz} method is not applicable in presence of an external potential.
Using LDA is generally good for the energy but misses the two-body correlations.
Instead quantum Monte Carlo methods can be efficiently used to tackle the problem.
Recently, lattice and path integral Monte Carlo algorithms were successfully used to study the properties of trapped bosons\cite{WeiRigol15}, trapped fermions with attraction\cite{BergerDrut15}, fermions in a box with periodic boundary\cite{Rammelmuller15}.
The properties of a balanced two-component Fermi gas in a harmonic trap were studied by means of the coupled-cluster method\cite{Grining15}.

In the following we study the energetic and structural properties of a trapped multi-component system of one dimensional fermions.

\section{The model Hamiltonian and parameters\label{Sec:H}}

We consider a multicomponent one-dimensional Fermi gas at $T=0$, trapped in a harmonic confinement of frequency $\omega$.
Inspired by the LENS experiment\cite{Pagano14}, we consider $N_c$ spin components of the same atomic species of mass $m$, with $N_p$ particles in each component, the total number of particles being equal to $N=N_c N_p$.
The system Hamiltonian is given by
\begin{equation}
H =
-\frac{\hbar^2}{2m}\sum_{\alpha=1}^{N_c}\sum_{i=1}^{N_p}\frac{\partial^2}{\partial^2 x_i^{\alpha}}
+g\sum_{\alpha<\beta}^{N_c}\sum_{i,j=1}^{N_p}\delta(x_i^{\alpha}-x_j^{\beta})
+\frac{m\omega^2}{2} \sum_{\alpha=1}^{N_c}\sum_{i=1}^{N_p}(x_i^{\alpha})^2,
\label{eq:H_1}
\end{equation}
where $g$ is the coupling constant.
In the LENS experiment $g$ was not changed.
However, in a more general case, its value can be fine-tuned by changing the magnetic field and exploiting the Feshbach and Olshanii\cite{Olshanii98} resonances.

In the LENS experiment\cite{Pagano14} the number of particles in each spin component was around $N_p = 20$.
Such a number is rather small, so it is questionable if all quantities can be precisely described within the local density approximation (LDA).
At the same time, this number is already larger than the system sizes which can be accessed with the direct diagonalization methods, as there the complexity grows exponentially with the system size.
On the contrary, quantum Monte Carlo methods work very efficiently with the system sizes of interest.

The harmonic confinement defines a characteristic length scale,
$a_{osc} = \sqrt{\hbar / (m \omega)}$
which we adopt as a unit of length.
We use the inter level spacing of a free confinement, $\hbar\omega$, as a unit of energy.
Another length scale is associated with the coupling constant, $g= -2\hbar^2/(ma_s)$, namely the $s$-wave scattering length $a_s$.
We remind that in one-dimension, the $s$-wave scattering length has a different sign compared to the usual three-dimensional case, that is $a_s$ is negative for a repulsive interaction, $g>0$.
Another peculiarity of a one-dimensional world is that the $s$-wave scattering length is inversely proportional to the coupling constant, for example, $a_s = 0$ corresponds to an infinite value of $g$.
The third characteristic length is the size of the system.
Within the local density approximation it is the Thomas-Fermi size, $R_{TF}$.
While in homogenous system the system properties are governed by a single dimensionless parameter, namely the gas parameter $na_s$, the presence of an external confinement requires, in general, an additional parameter.
Within the local density approximation, which cannot describe the Friedel oscillations and is expected to be applicable for large system sizes, $R_{RF}\gg a_{osc}$, the properties depend on the LDA parameter\cite{Astrakharchik04b,Astrakharchik2005LDA}
\begin{eqnarray}
\Lambda_{LDA} = N\frac{a_s^2}{a_{osc}^2}
\end{eqnarray}
At $Na_s^2/a_{osc}^2\rightarrow 0$ the interaction is infinitely strong and the ground-state energy of $N_c$-system becomes equal to the one of ideal one-component fermions.
In the opposite limit of $a_s^2/a_{osc}^2 \rightarrow \infty$ the interaction vanishes and the system behaves as $N_c$-component noninteracting Fermi gas.

\section{Methods\label{Sec:methods}}

We resort to the fixed-node diffusion Monte Carlo (FN-DMC) technique to find the ground-state properties of the system.
The proper fermionic symmetry of the wave function are imposed by using an antisymmetric trial wave function.
For a given nodal surface the FN-DMC provides the rigorous upper bound to the ground state energy.
If the nodal surface of the trial wave function is exact, the FN-DMC obtains the statistically exact ground-state properties of the system.
Importantly, the nodal surface in one dimension is known exactly as the fermionic wave function must vanish when any two fermions approach each other.

We chose the trial wave function as a product of determinants $S$ of a single-component Fermi gas and correlation terms which ensure Bethe-Peierls boundary condition\cite{Bethe35} between different species $\alpha$ and $\beta$,
$\partial \Psi / \partial (x_i^{\alpha}-x_j^{\beta})|_0  = - \Psi / a_s$:
\begin{equation}
\Psi_F({\bf R}) =
\prod_{\alpha=1}^{N_c} \det S({\bf R}_\alpha)
\times
\prod_{\alpha < \beta} \prod_{i=1}^{N_p} \prod_{j=1}^{N_p}||x_i^{\alpha}-x_j^{\beta}|-a_s|.
\label{eq:wf}
\end{equation}
Here ${\bf R}$ is the multidimensional vector which contains the coordinates $x_i^{\alpha}$ of all particles and
${\bf R}_\alpha$ contains all particles of component $\alpha$.
The Slater determinant in a harmonic trap is constructed from Hermite polynomials and has a special structure of the van der Monde determinant which can be explicitly evaluated\cite{Girardeau01} into a $N_p\times N_p$ pair product:
\begin{equation}
\det S({\bf R}_\alpha)
=
\prod_{i<j}^{N_p}| x_i^{\alpha}-x_j^{\alpha}|
\prod_{i=1}^{N_p} e^{-\gamma (x_i^{\alpha}/a_{osc})^2}\;,
\label{eq:wf:Slater}
\end{equation}
with $\gamma=1/2$ value corresponding to the ground-state wave function of a single-component Fermi gas.
Slater determinants~(\ref{eq:wf:Slater}) are antisymmetric with respect to exchange of same-component fermions and define the nodal surface of the total wave function~(\ref{eq:wf}).
We consider $\gamma$ as a variational parameter and optimize it value by minimizing the variational energy.
The nodal surface of the $\Psi_T({\bf R})$ is not affected by $\gamma$, but a proper choice of the variational parameter significantly reduces the statistical noise in the ground-state energy and in the correlation functions.

For comparison, we also consider a system of $N$ trapped bosons interacting via a contact potential.
In this case the trial wave function is chosen the following form:
\begin{equation}
\Psi_B({\bf R}) =
\prod_{i<j}^N||x_i-x_j| - a_s|
\prod_{\alpha=1}^{N}e^{-\gamma (x_i/a_{osc})^2}\;.
\label{eq:wf:bosons}
\end{equation}

\section{System energy and contact\label{Sec:energy}}

It is of a large interest to understand how the interplay between interactions and the quantum statistics affects the energetic properties of the system.
The release energy can be measured by suddenly removing the trapping potential and measuring the kinetic energy of the spreading gas.
Furthermore, the short-range interactions result in a very intrinsic relation between the interaction energy and the correlation functions, a phenomenon which is truly unique for dilute ultracold gases and is absent in condensed matter.
Namely, it was shown in 2008 by Shina Tan\cite{Tan2008a,Tan2008b,Tan2008c} that in three-dimensional Fermi gases many short-range properties of correlation function are related to a universal number, commonly referred as Tan's contact $C$, which, in turn, is related to the equation of state as
\begin{equation}
C=\frac{dE_0}{da_s}
\label{eq:Contact}
\end{equation}
In one dimension a similar relation was anticipated in 2003 by Maxim Olshanii and Vanja Dunko\cite{Olshanii03} (on 1D, see also Ref.~\cite{BarthZwerger11}).
Generally, the contact provides a number of universal relations which connect the short-range correlations to the thermodynamics of the whole many-body system.
The universality of the contact was experimentally verified in Bose\cite{WildJin12} and Fermi\cite{KuhnleVale10,SagiJin12} gases.

\begin{figure}[!ht]
\centering
\includegraphics[width=0.8\columnwidth]{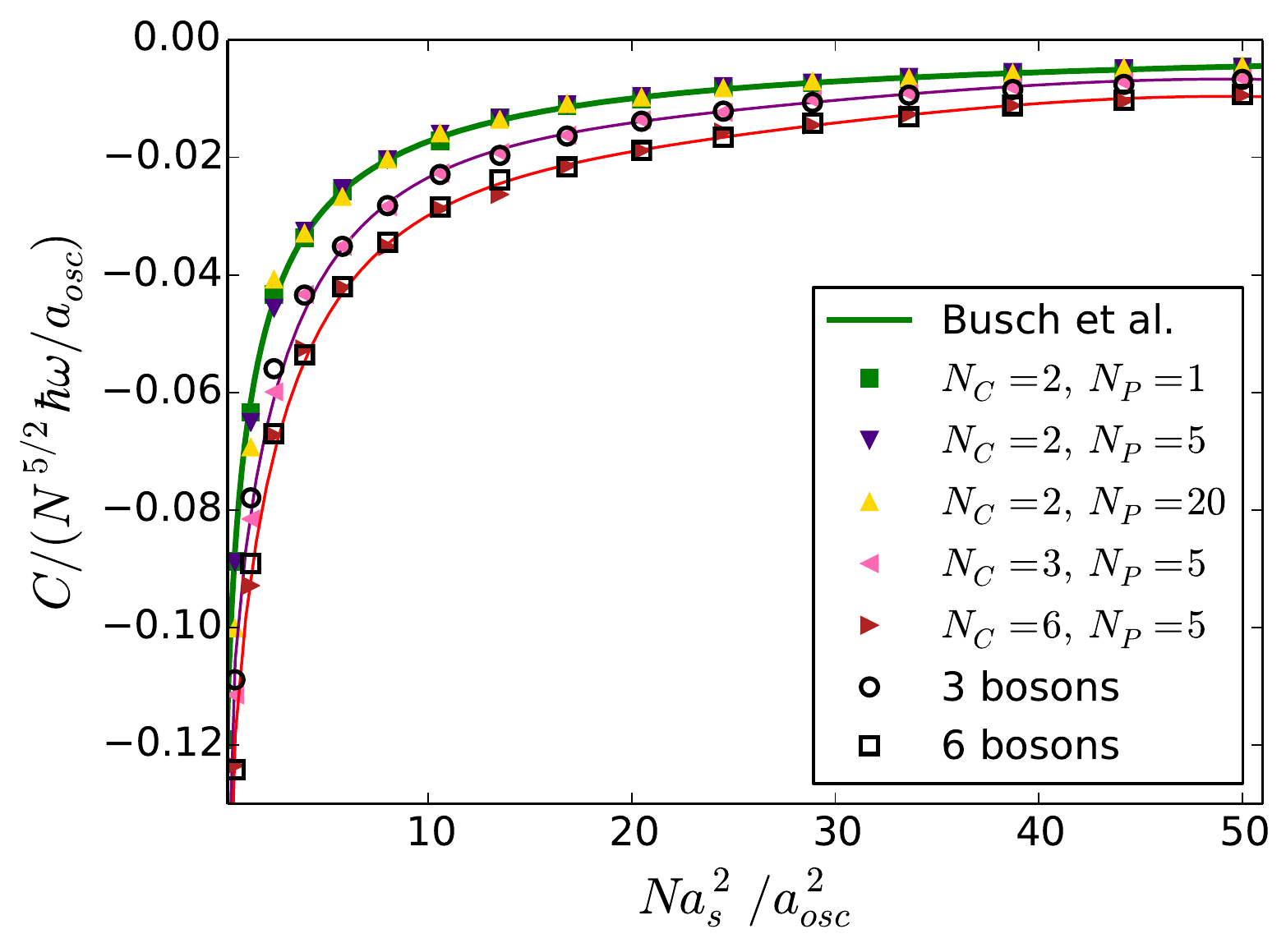}
\caption{The contact $C$ as a function of the interaction parameter $Na^2_s/a_{osc}^2$ for different number of components $N_c$ and with $N_p$ particles each.
The contact is obtained by differentiating the ground-state energy according to Eq.~(\ref{eq:Contact}).
Solid symbols, FN-DMC results for fermions;
open symbols, DMC results for bosons;
the green thick line, exact solution for $N_c=2$ and $N_p=1$ from \textit{Busch et al.} Ref.~\cite{Busch98};
thin lines,
polynomial fits to FN-DMC data.
The error bars are smaller than the symbol size. }
\label{fig:Contact}
\end{figure}

In order to obtain the FN-DMC value of contact $C$ we calculate derivative~(\ref{eq:Contact}) of the ground-state energy $E_0$ using finite difference.
The resulting dependence of the contact on parameter $Na_s^2/a_{osc}^2$ is reported in Fig.~\ref{fig:Contact}.
It is convenient to begin the analysis with the simplest case of a two-component system, $N_c = 2$.
For a single particle in each component, $N_p = 1$ (green solid squares), the many-body problem reduces to an exactly solvable two-body problem~\cite{Busch98} (green solid line).
A very good agreement between quantum Monte Carlo data and the analytic result verifies the used numerical approach.
For a single particle in each component, the quantum statistics is not yet important.
Instead, for $N_p = 5$ (purple down triangles) and $N_p = 20$ (yellow up triangles), particles in each component obey the Fermi-Dirac statistics.
Our numerical finding is that the rescaled contact $C/N^{5/2}$ has a negligible dependence on the number of particles in the component $N_p$ for the considered values of $Na_s^2/a_{osc}^2$.
From a practical point of view this allows to use the analytic two-particle result as a good approximation to the contact in a two-component system.

For fermions with three components, $N_c=3$ (pink left triangles), and six components, $N_c = 6$ (red right triangles), the absolute value of the contact is increased.
At the same time the qualitative dependence on $Na_s^2/a_{osc}^2$ remains the same.
Inspired by a good agreement between the contact in a two component fermions and two bosons, we also perform the simulations for the case of one-component bosons.
In this case $\Psi_T({\bf R}) = \Psi^{bos}_{J_1}({\bf R}) \Psi_G({\bf R})$, where in the first term $a_b = a$.
The DMC results for 3 and 6 bosons are shown in Fig.~\ref{fig:Contact} (black open circles and squares, correspondingly).
One can see that also in the case of a multi-component fermions the contact is similar to that of a gas of $N_c$ bosons in the considered range of parameters.

\begin{figure}[!ht]
\centering
\includegraphics[width=0.8\columnwidth]{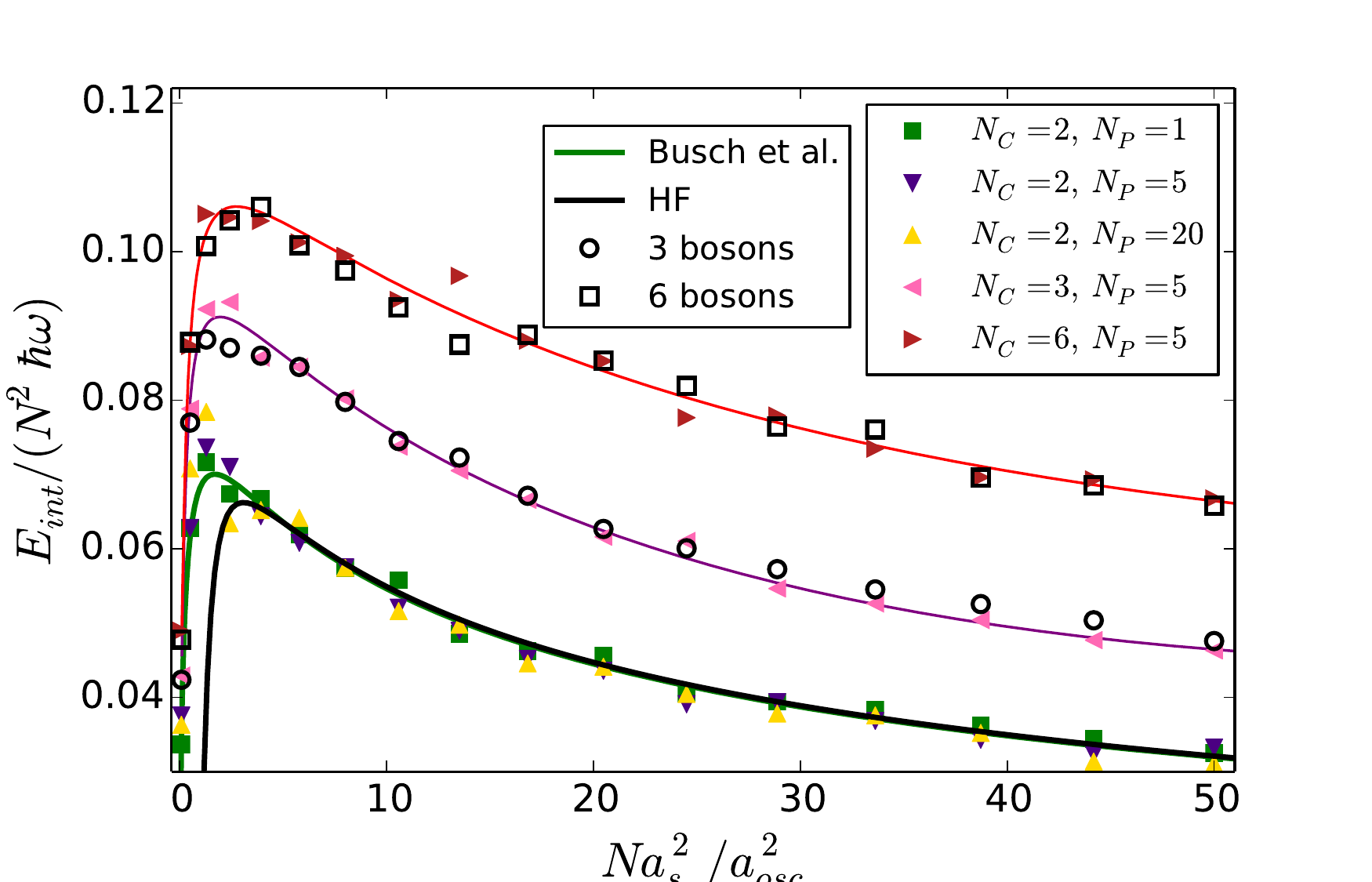}
\caption{The interaction energy $E_{int}$ as a function of the interaction parameter $Na_s^2/a_{osc}^2$ for different number of components $N_c$ and with $N_p$ particles each.
Solid symbols, FN-DMC results for fermions;
open symbols, DMC results for bosons;
green thick line, exact solution for $N_c=2$ and $N_p=1$ from \textit{Busch et al.} Ref.~\cite{Busch98};
black thick line, expansion~(\ref{eq:2particlesEint}) of the exact solution for $N_c=2$ and $N_p=1$  in the limit of weak interactions; 
thin lines,
polynomial fits to FN-DMC data.
The error bars are smaller than the symbol size.
}
\label{fig:Eint}
\end{figure}

The interaction energy,
$E_{int} = \langle g\sum_{\alpha<\beta}^{N_c}\sum_{i,j=1}^{N_p}\delta(x_i^{\alpha}-x_j^{\beta}) \rangle$, can be calculated by using Hellmann-Feynman theorem as $E_{int} = -a_sdE_0/da_s$.
In other words, there is a close relation between the contact~(\ref{eq:Contact}) and the interaction energy for a contact potential,
\begin{equation}
E_{int} = -aC\;.
\label{eq:Eint}
\end{equation}
We show the interaction energy in Fig.~$\ref{fig:Eint}$.
The interaction energy vanishes in the limit of non-interacting species, $g\to 0$, which corresponds to $Na_{s}/a_{ho}\to\infty$. In the opposite limit of very strong interspecies interactions, $g\to +\infty$, there is a node in the wave function when particles approach each other (refer to Eq.~(\ref{eq:wf}) with $a_s = 0$).
As a result the interaction energy vanishes also in the limit $Na_{s}/a_{ho}\to 0$.
The maximum in the interaction energy is observed for $Na_{s}/a_{ho} \approx 1$.

Similarly to the contact, we find that the most important effect is the number of components $N_c$ and not the number of particles in each of them, at least for the considered parameter range.

\section{Correlation functions}

In one-dimensional geometry there exists an unusual relation between the interactions and statistics, which is absent in higher dimensions.
By tuning the interspecies interaction strength it is possible to change the correlations from uncorrelated ``ideal boson-like'' case to the ``ideal fermion-like`` case.

According to Girardeau's mapping\cite{Girardeau60}, infinite repulsion between one-dimensional particles mimics the Pauli principle and the absolute value of the wave function is the same.
This holds true both for a system of one-component bosons (Tonks-Girardeau gas) but also a system of two-component Fermions.
For the case $N_c=N_p=1$ the fermionization of two distinguishable fermions was experimentally demonstrated in Selim Jochim's group\cite{Zurn12}.
From the Girardeau's mapping it immediately follows that the diagonal properties, which depend on the square of the absolute value of the wave function, are the same in both system.
Instead, off-diagonal properties are different.
From that it is highly instructive to analyze the evolution of both one- and two-body correlation functions.
We consider the density profile $n(x)$ and the density-density correlation function $g(x)$, which obey the Girardeau's mapping, and experimentally relevant momentum distribution $n(k)$, for which the mapping does not apply.

\subsection{Density profile}

\begin{figure}[!ht]
\centering
\includegraphics[width=0.8\columnwidth]{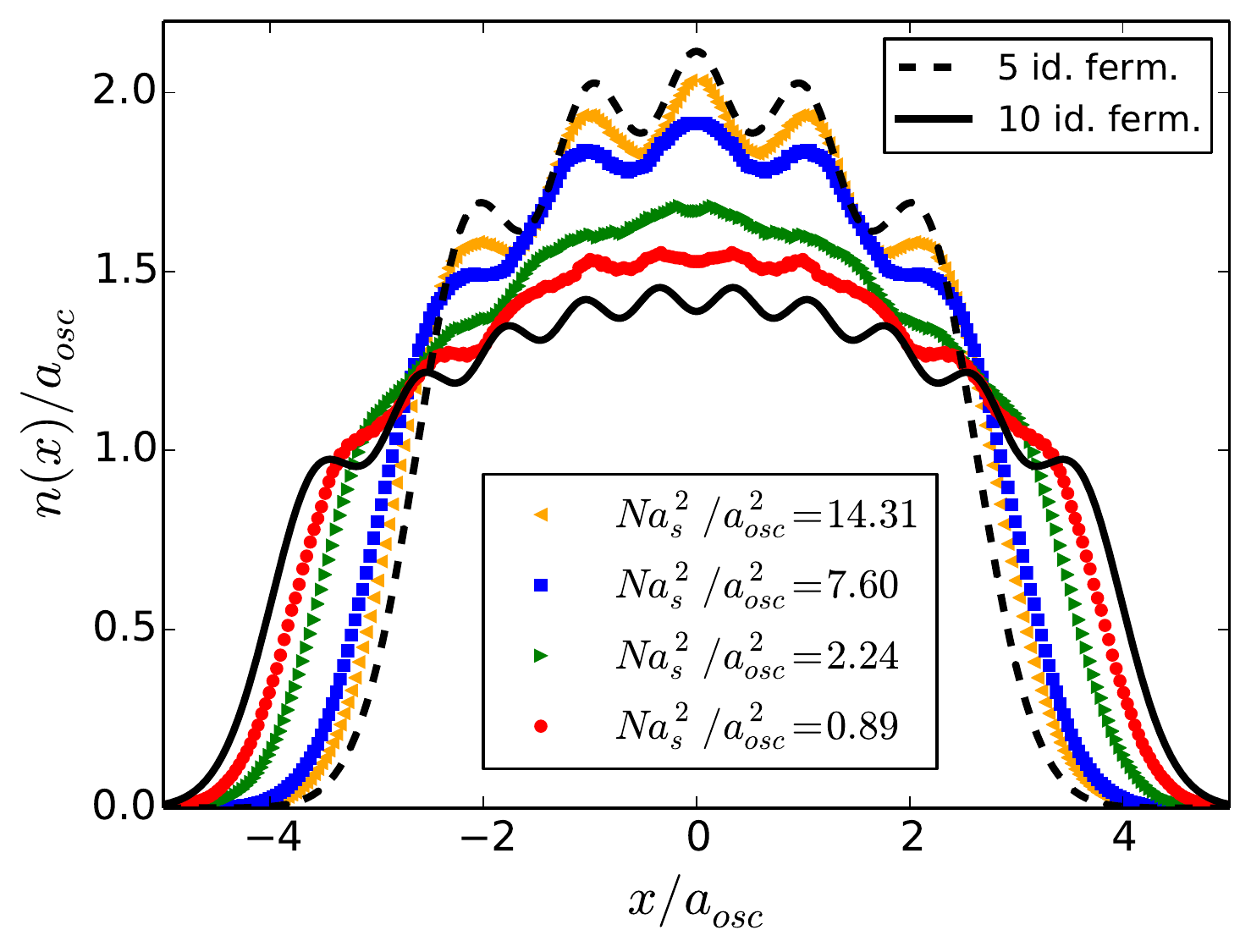}
\caption{The density distribution $n(x)$ for a two-component Fermi gas $N_c = 2$ at different values of parameter $Na_s^2/a_{osc}^2 = 14.31; 7.60; 2.24; 0.89$ (decreasing the density in the center $n(x=0)$).
The dotted (solid) black line shows the density distribution for 5 (10) ideal spin-polarized fermions.}
\label{fig:dens_distr}
\end{figure}

We calculate the density profile $n(x)$ in a two-component Fermi gas, $N_c=2$, with $N_p = 5$ particles in each component, as a function of the interaction parameter $Na_s^2/a_{osc}^2$ and show it in Fig.~\ref{fig:dens_distr}).
For no interaction between two species, $Na_s^2/a_{osc}^2 \to \infty$, the density profile is given by that of a spin-polarized Fermi gas with $N_p$ particles (dashed line in Fig.~\ref{fig:dens_distr}).
In this case, the interaction energy $E_{int}=0$ (refer to Fig.~\ref{fig:Eint}).
As the interaction strength is increased, the interaction energy increases and the system size becomes larger.
Stronger repulsion pushes the particles to the edges of the trap, making the density in the center drop down.
For infinite interaction strength, $Na_s^2/a_{osc}^2 \to 0$, Girardeau's arguments predict that wave function can be mapped to that of an ideal single-component Fermi gas of $N_cN_p$ particles.
Indeed, we see that in this limit the density profile is that of an ideal Fermi gas with $2N_p$ particles (solid line).
In between we see a smooth crossover from one regime to the other.
In other words we observe a fermionization of two fermion species, similarly to experimentally observed $N_p=1$ case\cite{Zurn12} (see also Ref.~\cite{Fang09} for Bose-Fermi mixtures).

It is important to note that the Friedel oscillations, clearly seen in Fig.~\ref{fig:dens_distr} are completely missed within local density approximation and cannot be predicted using Bethe solutions for homogeneous gas.
Furthermore, here we observe an interesting parity effect.
When interspecies interaction is absent, there is a maximum at $n(x=0)$.
For infinite interspecies interaction, there is a minimum at $n(x=0)$.
In addition we see a doubling in the number of peaks which sometimes is interpreted as a transition from $2k_F$-Friedel to $4k_F$-Wigner oscillations~\cite{Xianlong12}.

On the other hand, even if LDA misses the shell structure, it is still capable to predict the overall shape and the cloud size.
As well a polytropic approximation to the equation of state permits to understand some basic features analytically.
In the non-interacting limit, $Na_s^2/a_{ho}^2 \to\infty$, the density profile is a semicircle,
$n(x) \propto \sqrt(\mu - m\omega^2x^2/2)$  with $\mu = N_p\hbar\omega$.
In the strongly-interacting limit, $Na_s^2/a_{ho}^2 \to 0$, the density profile is again a semicircle but of a larger size, $\mu = 2N_p\hbar\omega$.
In the limit of large number of components, $N=N_c$ and $N_p=1$, the density profile is a semicircle for infinite interaction, $\mu=N\hbar\omega$, is an inverted parabola for weak interactions, $n(x) \propto (\mu - m\omega^2x^2/2)$;
and is a free-harmonic oscillator Gaussian profile for zero interactions (not captured by LDA).

\subsection{Pair correlation function}

The pair correlation function $g(x)$ is proportional to the probability of finding a pair of particles separated by a distance $x$.
It is a diagonal quantity and is a subject to Girardeau's mapping, so we expect to see fermionization for strong repulsive interactions.
In a trapped system the long-range decay zero comes from vanishing particle density at the edges of the trap and is eventually a one-body effect.
From the quantum-correlation perspective the most interesting behaviour is that of short and intermediate distances.

For the intra-component correlation function the Pauli exclusion principle implies that $g^{\uparrow\uparrow}(x=0) = 0$.
Instead, for inter-component correlation function $g^{\uparrow\downarrow}(x)$ the value at the contact, $x=0$, is related to $C$ using the following relations:
\begin{eqnarray}
\frac{C}{N} &=& \frac{g(0)}{a_s^2}\;.
\label{eq:g(0)}
\end{eqnarray}

\begin{figure}[!ht]
\centering
\includegraphics[width=0.8\columnwidth]{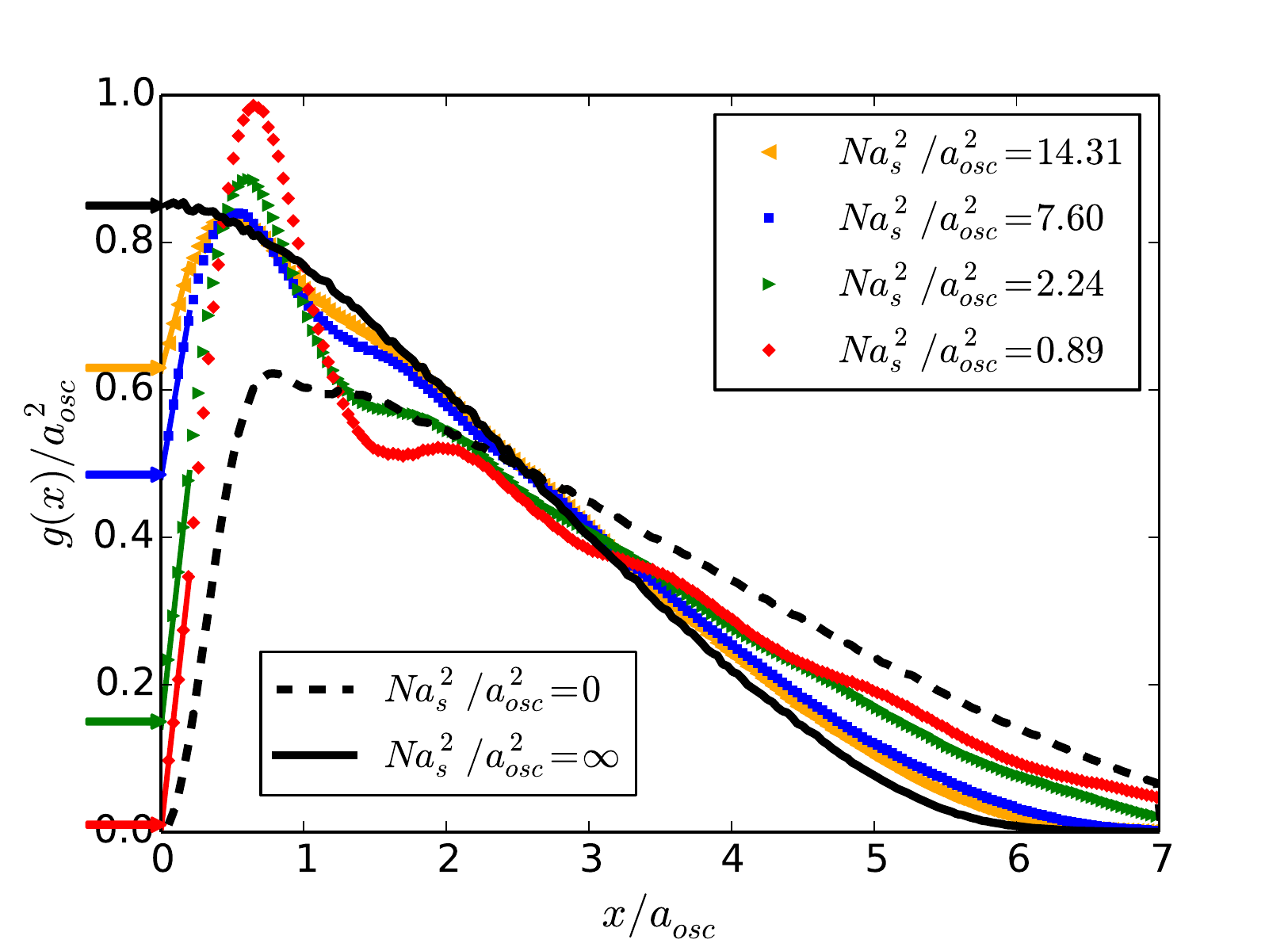}
\caption{The pair-correlation function $g^{\uparrow\downarrow}(x)$ in a two-component Fermi system, $N_c = 2$, for different values of parameter $Na_s^2/a_{osc}^2 = 14.31; 7.60; 2.24; 0.89$ (decreasing the value at $g(x=0)$).
The horizontal arrows mark the value of the contact obtained from the numerical derivative of the FNDMC ground state energy.
}
\label{fig:gr}
\end{figure}

In Fig.~\ref{fig:gr} we show the inter-component pair-correlation function $g^{\uparrow\downarrow}(x)$ in a two-component system. One can observe how the decrease of the interaction parameter $Na_s^2/a_{osc}^2$ (which means the increase of the interaction strength) the value $g^{\uparrow\downarrow}(x=0)$ decreases until it reaches the zero value corresponding to a fully fermionized system.
The horizontal arrows in Fig.~\ref{fig:gr} shows $g(0)$ at given $Na_s^2/a_{osc}^2$ calculated from the FN-DMC energy through the the contact, see Eq.~\ref{eq:Contact}.
We observe a good agreement between these two methods of the calculation of the contact.

\subsection{Momentum distribution}

An important experimentally observable quantity is the momentum distribution $n(k)$.
Being an off-diagonal quantity, it will not fermionize in the limit of infinitely-strong repulsion and will be essentially different from $n(k)$ of an ideal Fermi gas.
Also, the use of the Bethe {\it ansatz} methods is very limited in calculation of $n(k)$.

\begin{figure}[ht]
\centering
\subfigure[]{
\centering
\includegraphics[width=0.8\columnwidth]{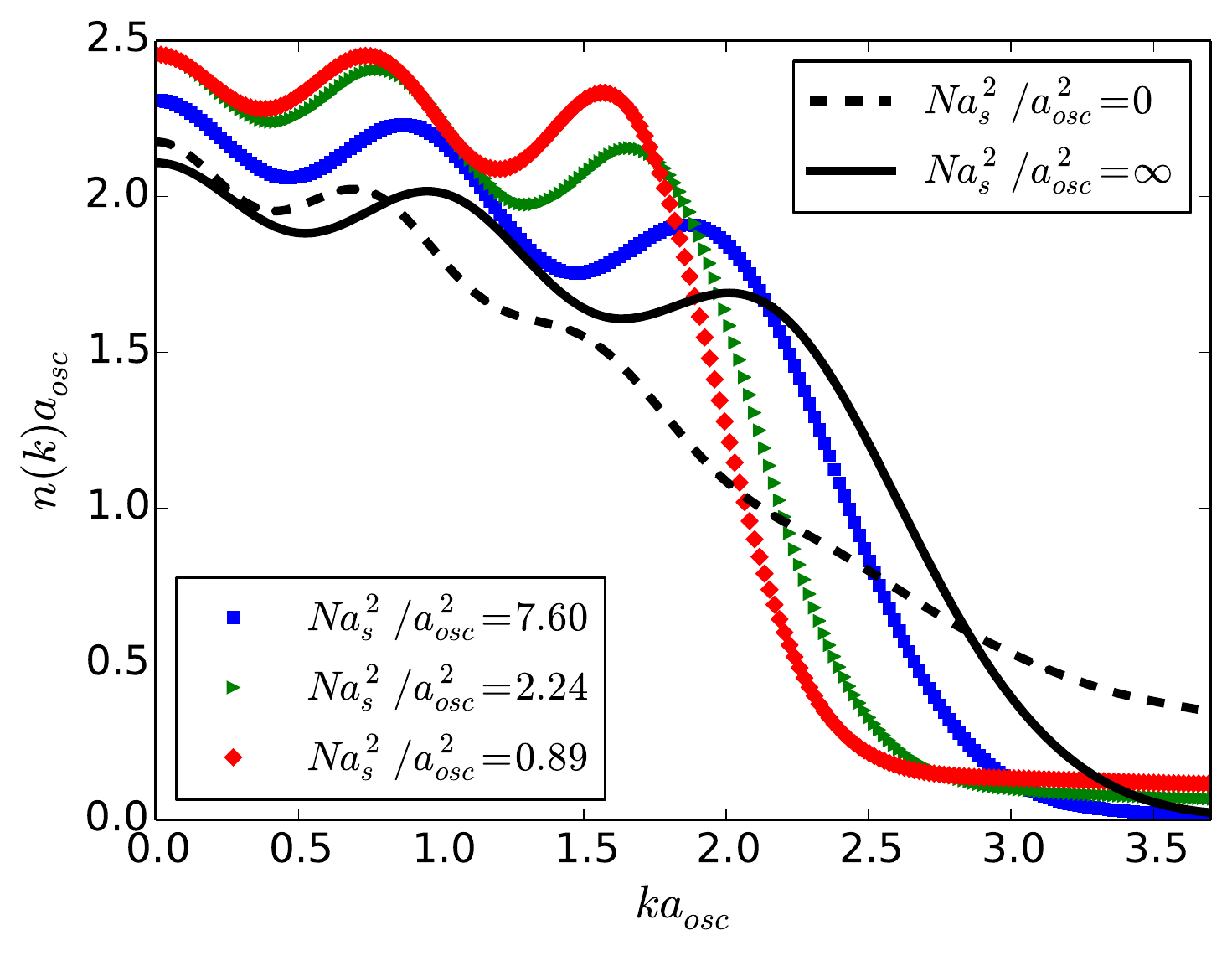}
\label{fig:subfig1}
}
\subfigure[]{
\includegraphics[width=0.8\columnwidth]{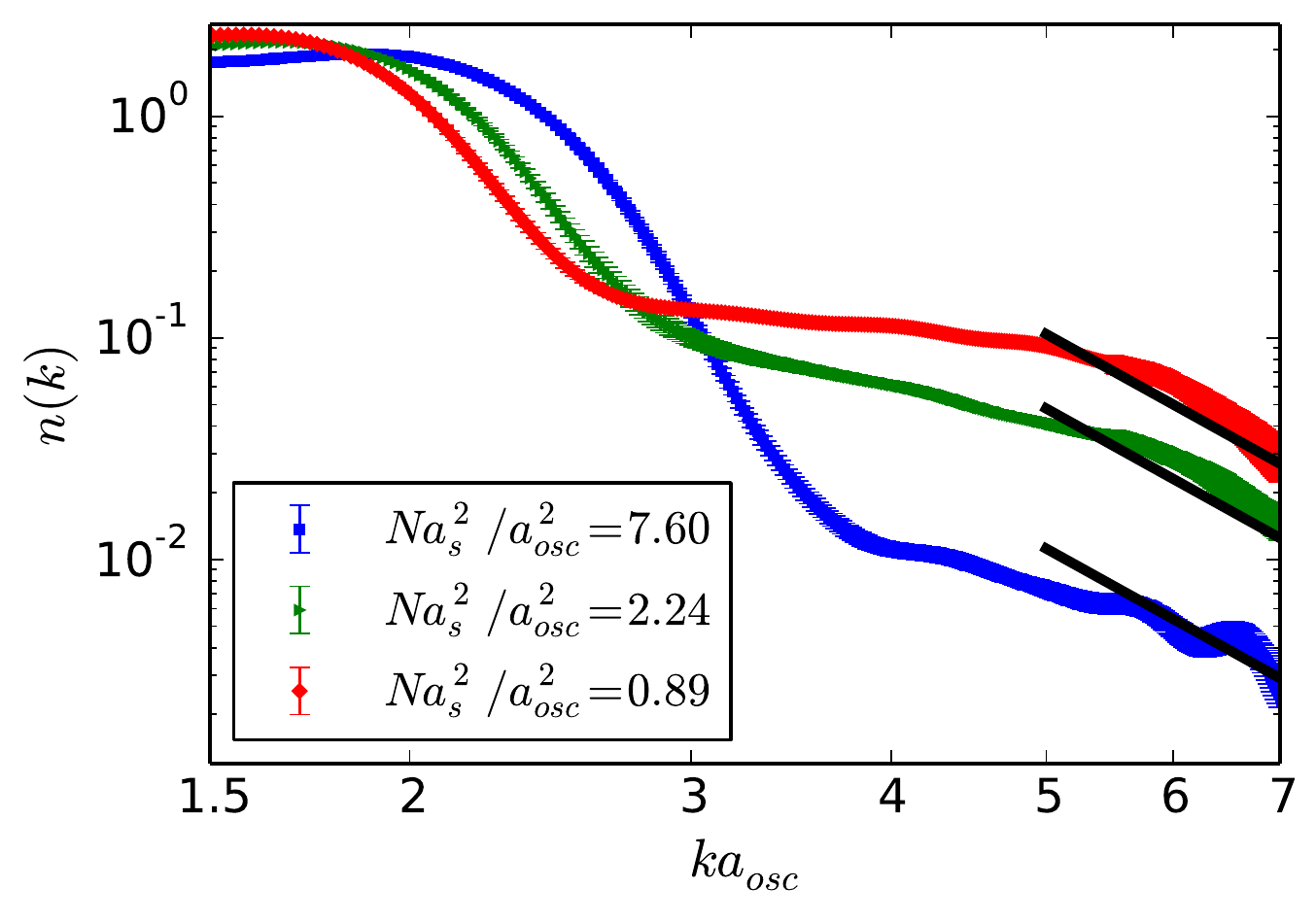}
\label{fig:subfig2}
}
\caption[Optional caption for list of figures]{
The momentum distribution $n(k)$ in a two-component Fermi system, $N_c = 2$, for different values of parameter $Na_s^2/a_{osc}^2 = 7.60; 2.24; 0.89$ (decreasing the height of the last peak),
(\subref{fig:subfig1}) on linear scale
(\subref{fig:subfig2}) on log-log scale.
The black lines at large $k$ shows the function $C/k^4$, where $C$ is the value of the contact obtained from the derivative of the energy, Eq.~(\ref{fig:Contact}). }
\label{fig:mom_distr}
\end{figure}

We show the momentum distribution of a two component system in Fig.~\ref{fig:mom_distr}.
One observes that with the decrease of the parameter $Na_s^2/a_{ho}^2$ (increasing the interactions) the amplitude of $n(k)$ increases and the distribution becomes narrower.
This latter effect is opposite to the particle distribution becoming wider in the real space, see Fig.~\ref{fig:dens_distr}.
The value of the Tan's contact can be extracted from the high-momentum tail of $n(k)$ according to
\begin{equation}
n(k) = C/k^4, \quad |k|\to\infty
\label{eq:nk}
\end{equation}
To see that we plot the same results on a double logarithmic scale in Fig.~\ref{fig:subfig2}.
The black lines at large $k$ show the function $C/k^4$, where $C$ is the value of the contact obtained from the numerical derivative of the energy.
We find that asymptotic regime is reached for momenta larger than $k a_{osc} \approx 5$.

\section{Conclusions}

To conclude, we studied the ground-state properties of a one-dimensional multicomponent Fermi gas in presence of a harmonic confinement ($N_c$ components with $N_p$ particles each).
Interspecies interaction is considered to be of a contact form, $g \delta(x^\uparrow - x^\downarrow)$

We used fixed-node diffusion Monte Carlo (FN-DMC) method to calculate the energy and correlation functions.
A unusual feature of one-dimensional geometry is that there is an interplay between quantum statistics and contact-interaction.
According to Girardeau's mapping\cite{Girardeau60}, the wave function in the limit of infinite repulsion, $g\to+\infty$, can be mapped to the wave function of an ideal Fermi gas.
In this regime, the multicomponent gas {\it fermionizes} having the same energy and diagonal properties as a single-component ideal Fermi gas of $N = N_c N_p$ particles.
The fermionization is observed in the energy, the density profile, the pair correlation function but not in the momentum distribution.

We calculate the Tan's contact $C$ for different number of particles $N_p$ in each of $N_c$ components.
We find that for a given number of components $N_c$, the $C / N^{5/2}$ has an almost universal dependence on parameter $N a_s^2/a^2_{osc}$ for the considered interaction and number of particles.
Importantly, we discover that in our case the contact can be well approximated by considering only one particle in each component $N_p=1$.
For a two-component Fermi gas $N_c = 2$ in harmonic confinement this allows to use analytic result of Busch et al.\cite{Busch98}. Also we check numerically that for $N_c = 3$ and $N_c = 6$ components the value of the Tan's contact is close to that of a Bose system with 3 and 6 particles.
We verify that the Tan's contact, calculated from the derivative of the ground-state energy~(\ref{eq:Contact}), as well provides the description of the interaction energy~(\ref{eq:Eint}), value of the pair-correlation function at zero~(\ref{eq:g(0)}), and large-momentum tail of the momentum distribution~(\ref{eq:nk}).

\section*{Acknowledgements}
NM would like to acknowledge the Nanosciences Foundation of Grenoble for financial support.
GEA acknowledges partial financial support from the MICINN (Spain) Grant No.~FIS2014-56257-C2-1-P.
The Barcelona Supercomputing Center (The Spanish National Supercomputing Center -- Centro Nacional de Supercomputaci\'on) is
acknowledged for the provided computational facilities.

\section*{Appendix: contact for weak interaction, LDA}

The limit of weak interaction can be analyzed assuming ideal Fermi gas density profile for each component and taking into account the interaction potential perturbatively.
For energetic calculations it is sufficient to assume the LDA shape, which for a single component provides the correct energy even for $N_p = 1$
\begin{equation}
n(x) = \frac{1}{\pi a_{osc}} \sqrt{2N_p - (x/a_{osc})^2}
\label{eq:LDA}
\end{equation}
The interaction energy is
\begin{equation}
E_{int}
= g\sum_{\alpha<\beta}^{N_c}\int dx_\alpha \int dx_\beta n(x_\alpha)n(x_\beta) \delta(x_i^{\alpha}-x_j^{\beta})
= g \frac{N_c(N_c-1)}{2}
\int n^2(x) dx
\label{eq:Eint:perturbative}
\end{equation}
Considering an unperturbed LDA density profile~(\ref{eq:LDA}) we get from~(\ref{eq:Eint:perturbative}) the interaction energy
\begin{equation}
E_{int} = \frac{8 \sqrt{2}}{3\pi^2} (N_c-1)N_c N_p^{3/2} \frac{\hbar^2}{m a_{osc} |a_s|}
\label{eq:Eint:LDA}
\end{equation}
and the contact
\begin{equation}
C = -\frac{8 \sqrt{2}}{3\pi^2} (N_c-1)N_c N_p^{3/2} \frac{\hbar^2}{m a_{osc} a_s^2}\;.
\label{eq:contact:LDA}
\end{equation}
We note that Eqs.~(\ref{eq:Eint:LDA}-\ref{eq:contact:LDA}) are derived assuming weak interaction, $a_s\to -\infty$ and a large number of particles $N_p$ in each component.
A priori, it is not obvious if $N_p=1$,
$E_{int}/(\hbar\omega) = -0.77 a_{osc}/{a_s}$, expression~(\ref{eq:Eint:LDA}) can be used at all.
Nevertheless, a comparison with the exact expression~(\ref{eq:2particlesEint}) is less than 5\% of the exact coefficient in the asymptotic expansion. 
The major difference between multiple-components and two-particle cases is the presence of a combinatoric $N_C(N_C-2)/2$ term in Eq.~(\ref{eq:Eint:LDA}).

\section*{Appendix: contact for weak interaction, two particles}

The problem of two particles in a harmonic trap can be solved analytically\cite{Busch98}.
The harmonic confinement permits to separate the problem into center of mass (CM) and relative motion (rel).
The ground-state energy of the CM motion is $E_{CM} = \hbar\omega/2$.
The energy of the relative motion $E_{rel}$ is defined as a solution to the following equation (harmonic oscillator units are used)
\begin{equation}
\sqrt{2} \frac{\Gamma[-E_{rel}/2 + 3/4]}{\Gamma[-E_{rel}/2 + 1/4]} = \frac{1}{a_s}\;.
\label{eq:2particlesE}
\end{equation}
For the same value of the $s$-wave scattering length $a_s$ Eq.~(\ref{eq:2particlesE}) permits multiple solutions for the energy, which correspond to multiple level structure.
For a weakly interacting case, $a_s\to-\infty$, the energy of the relative motion is close to $\hbar\omega/2$ value and Gamma functions in Eq.~(\ref{eq:2particlesE}) can be expanded into series around this point.
The resulting total energy $E = E_{rel} + E_{CM}$ is 
\begin{equation}
E
=
1 
- \sqrt{\frac{2}{\pi}}\frac{1}{a_s}
- \frac{\ln 4}{\pi a_s^2}
-\frac{0.1572}{a_s^3}
-\frac{0.0196}{a_s^4}
+\frac{0.0129}{a_s^5}
\end{equation}
The interaction energy can be obtained using Hellmann-Feynman theorem
\begin{eqnarray}
\label{eq:2particlesEint}
E_{int} &=&
- \sqrt{\frac{2}{\pi}} \frac{1}{a_s}
- \frac{2\ln 4}{\pi}\frac{a_{osc}^2}{a_s^2} + ...\\
&=& - \frac{0.797885}{a_s} 
- \frac{0.441271}{a_s^2}
+ \frac{0.471548}{a_s^3} 
-\frac{0.0784608}{a_s^4} 
+ \frac{0.0645712}{a_s^5}
\nonumber
\end{eqnarray}
And, finally, the contact is 
\begin{equation}
C = \sqrt{\frac{2}{\pi}} \frac{\hbar^2}{m a_{osc} a_s^2} 
+ \frac{2\ln 4}{\pi} \frac{\hbar^2}{m a_s^3} 
\label{eq:2particlescontact}
\end{equation}

\section*{Bibliography}

\bibliographystyle{ieeetr}

\end{document}